\documentclass[aps,prb,reprint,
               superscriptaddress,
               floatfix,
               amsmath,amssymb,
               longbibliography]{revtex4-2}
\AtBeginDocument{\hypersetup{colorlinks=true,linkcolor=black,citecolor=black,urlcolor=black}}

\usepackage{graphicx}
\usepackage{dcolumn}
\usepackage{bm}
\usepackage{xcolor}
\usepackage{pifont}
\usepackage{booktabs}
\usepackage{multirow}
\usepackage{array}
\usepackage{orcidlink}
\usepackage{placeins}

\begin{document}

\title{Role of Chalcogen-$p$ States in the Optical Response of Bulk PtSe$_2$ and PtTe$_2$}

\author{Ji Eun Lee\orcidlink{0000-0001-8947-826X}}
\email{Contact author: jieun.lee@pi1.uni-stuttgart.de}
\affiliation{1.~Physikalisches Institut, Universit\"at Stuttgart, Pfaffenwaldring 57,
70569 Stuttgart, Germany}

\author{R. Mathew Roy\orcidlink{0009-0003-2842-9044}}
\affiliation{1.~Physikalisches Institut, Universit\"at Stuttgart, Pfaffenwaldring 57,
70569 Stuttgart, Germany}

\author{P. Yanda\orcidlink{0000-0001-9798-1236}}
\affiliation{Max Planck Institute for Chemical Physics of Solids, 01187 Dresden, Germany}

\author{C. Shekhar\orcidlink{0000-0002-3330-0400}}
\affiliation{Max Planck Institute for Chemical Physics of Solids, 01187 Dresden, Germany}

\author{C. Felser\orcidlink{0000-0002-8200-2063}}
\affiliation{Max Planck Institute for Chemical Physics of Solids, 01187 Dresden, Germany}

\author{M. Dressel\orcidlink{0000-0003-1907-052X}}
\affiliation{1.~Physikalisches Institut, Universit\"at Stuttgart, Pfaffenwaldring 57,
70569 Stuttgart, Germany}

\date{\today}

\begin{abstract}
The type-II Dirac semimetals PtSe$_2$ and PtTe$_2$ are isostructural and isovalent layered compounds but exhibit distinct electronic properties. We measure the in-plane infrared reflectivity of bulk crystals between 10 and 300~K and compare the interband contribution to the optical conductivity with band-resolved density-functional calculations. At low temperature, the Drude spectral weight of PtTe$_2$ is an order of magnitude larger than that of PtSe$_2$. PtTe$_2$ also shows a broad mid-infrared absorption continuum, whereas the absorption of PtSe$_2$ remains weak. The calculations attribute this continuum to transitions from a predominantly chalcogen-$p$ band into a Pt-$d$ band. In PtTe$_2$, the chalcogen-$p$ band lies higher in energy, which lowers the onset of these transitions and increases the semimetallic band overlap, in line with the larger Drude spectral weight. The type-II Dirac points lie far below the Fermi level, so that transitions between the bands forming the Dirac cones do not contribute appreciably to the interband conductivity.
\end{abstract}

\maketitle

\begin{figure*}[!htb]
\centering
\includegraphics[width=1.0\textwidth]{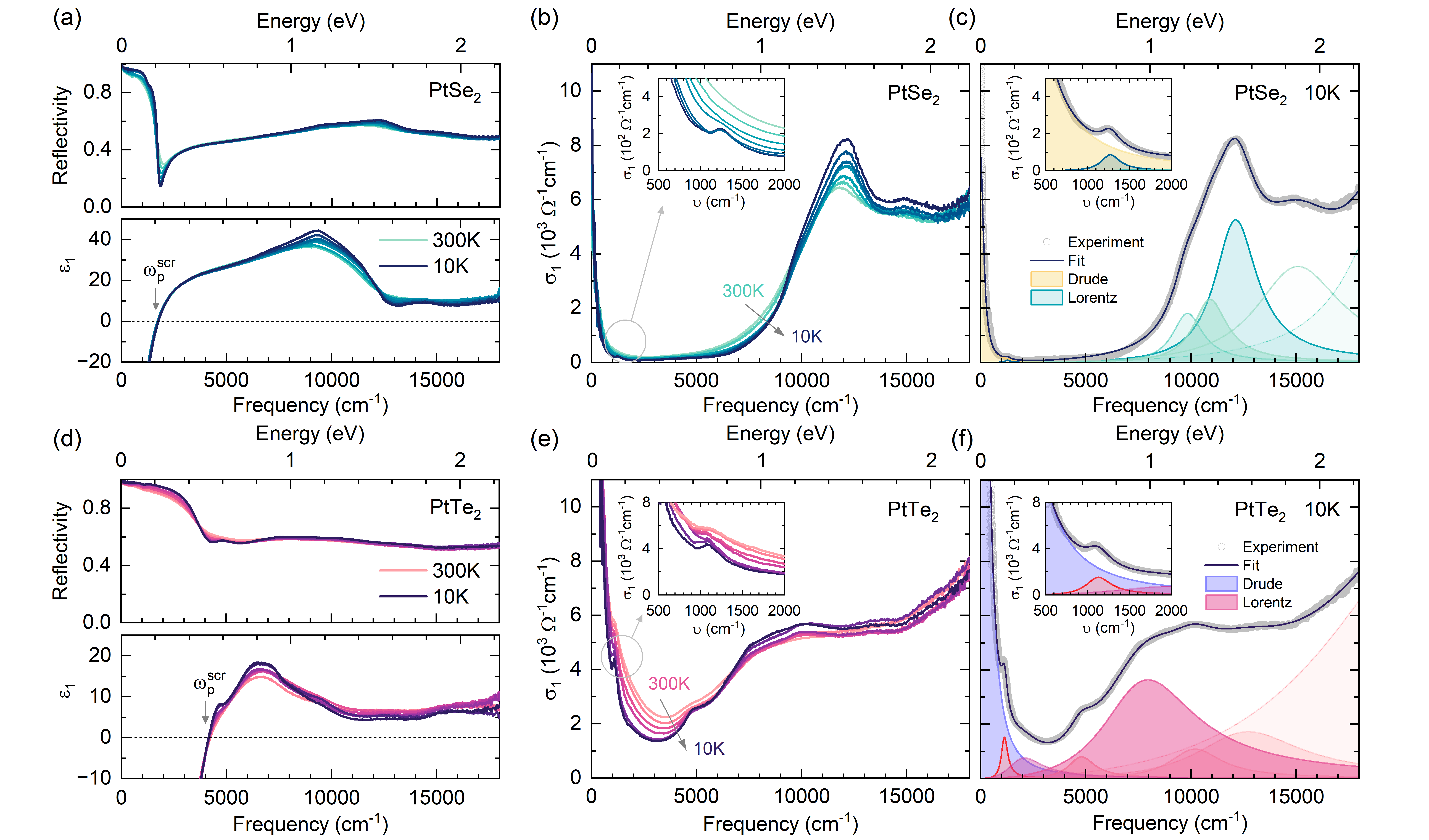}
\caption{(a) Temperature-dependent in-plane reflectivity (top) and real part of the dielectric function $\varepsilon_1(\omega)$ (bottom) of PtSe$_2$ at $T=10$, 50, 100, 150, 200, 250, and 300~K. The arrow marks the zero crossing of $\varepsilon_1(\omega)$, which corresponds to the screened plasma frequency $\omega_p^{\mathrm{scr}}$. (b) Real part of the in-plane optical conductivity $\sigma_1(\omega)$. The inset shows the circled region around the low-energy peak on an enlarged scale. (c) Decomposed optical conductivity at 10~K, consisting of a Drude contribution and several Lorentzians describing the interband transitions, with the total fit. The inset shows the Drude contribution and the Lorentzian of the low-energy peak on an enlarged scale. (d)--(f) Same as (a)--(c) for PtTe$_2$.}
\label{Fig1}
\end{figure*}
\begin{figure*}[!htb]
\centering
\includegraphics[width=1.0\textwidth]{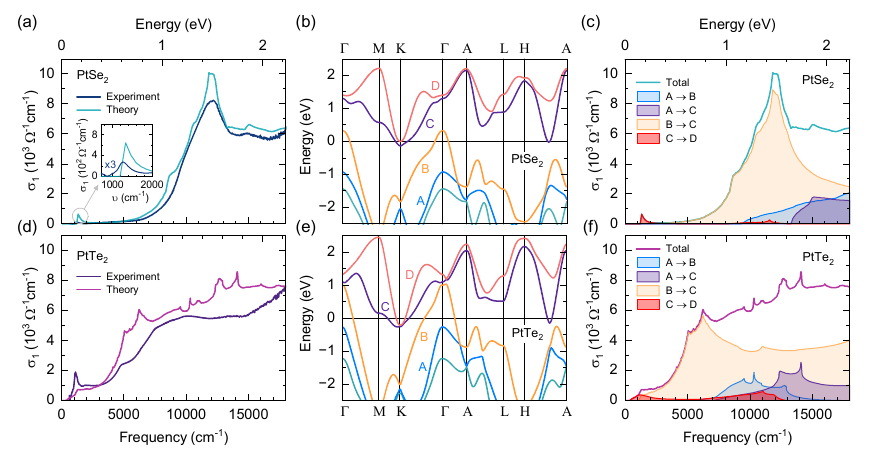}
\caption{(a) Experimental in-plane interband optical conductivity of PtSe$_2$ at 10~K, obtained by subtracting the Drude contribution from the spectrum, compared with the DFT-based interband optical conductivity. The energy axes of the calculated spectra in (a) and (c) are multiplied by a factor of 1.21. The inset shows the low-energy peak on an enlarged scale, with the experimental spectrum multiplied by a factor of 3. (b) Calculated band structure of PtSe$_2$ with the bands labeled A--D. (c)~Band-resolved interband optical conductivity, showing the contributions from transitions between the bands labeled in (b). (d)--(f) Same as (a)--(c) for PtTe$_2$, without energy scaling.}
\label{Fig2}
\end{figure*}

\section{Introduction}
The platinum dichalcogenides Pt$X_2$ ($X$ = S, Se, Te) crystallize in the layered CdI$_2$-type $1T$ structure with space group $P\bar{3}m1$~\cite{guo1986electronic}. Despite the common structure, their electronic properties depend strongly on the chalcogen. Bulk PtS$_2$ is a semiconductor, whereas PtSe$_2$ and PtTe$_2$ are semimetals~\cite{villaos2019thickness}. Electronic-structure calculations attribute this difference to the upward shift of the chalcogen-$p$ levels toward the Pt-$d$ levels and the accompanying stronger antibonding interaction between the chalcogen-$p_z$ orbitals of adjacent layers~\cite{dai2003trends}. In addition, PtSe$_2$ and PtTe$_2$ host type-II Dirac fermions, which were predicted by first-principles calculations~\cite{huang2016typeII} and subsequently observed by angle-resolved photoemission spectroscopy (ARPES)~\cite{zhang2017ptse2,yan2017ptte2,pelayo2024threedimensional}.

Bulk PtSe$_2$ and PtTe$_2$ show metallic resistivity, which is substantially lower in PtTe$_2$ than in PtSe$_2$~\cite{zhang2017ptse2}. Quantum-oscillation and magnetotransport studies revealed a multiband electronic structure with several electron and hole pockets near the Fermi level $E_{\mathrm{F}}$~\cite{yang2018quantum,zheng2018detailed,pavlosiuk2018galvanomagnetic}. Beyond these bulk properties, PtSe$_2$ and PtTe$_2$ exhibit a strong thickness dependence of their electronic structure~\cite{villaos2019thickness,kempt2020two}. In thin layers, quantum confinement and weaker interlayer coupling open a band gap~\cite{miro2014two,kandemir2018structural,ciarrocchi2018thickness,ansari2019quantum}, and semiconducting behavior has been reported for monolayer and bilayer PtSe$_2$ and for monolayer PtTe$_2$~\cite{li2021layer}. Thin films of PtSe$_2$ and PtTe$_2$ have been explored for infrared photodetection~\cite{yu2018atomically,zeng2020van} and terahertz detection and emission~\cite{xu2019ptte2,cheng2023giant,chen2025giant}.

Optical studies of PtSe$_2$ have focused on thin films and exfoliated flakes. Ellipsometry and absorption measurements on films and flakes a few nanometers thick characterized the interband response from the near-infrared to the ultraviolet range~\cite{xie2019optical,gulo2020temperature} and attributed the observed transitions to states derived from Pt-$d$ and Se-$p$ orbitals~\cite{tharrault2025optical}. Terahertz spectroscopy and ellipsometry followed the thickness-driven semiconductor-to-semimetal transition~\cite{ji2023probing,micica2025bandgaps}, while infrared absorbance and mid-infrared photocurrent measurements were performed on polycrystalline films~\cite{parhizkar2022waveguide,gerlei2026strain}. For PtTe$_2$, electron energy-loss spectroscopy (EELS) on bulk crystals revealed a plasmon near 0.5~eV, which was interpreted as a three-dimensional Dirac plasmon on the basis of an effective Dirac-cone model~\cite{politano2018dirac,ghosh2019broadband}. Plasmon polaritons in PtTe$_2$ films and nanoribbons were further studied by infrared spectroscopy and nanoimaging and were also interpreted in terms of Dirac fermions~\cite{macis2024terahertz,hu2020infrared}. Despite these studies, a comparison of the optical conductivity of bulk PtSe$_2$ and PtTe$_2$ is still lacking.

In this work, we present a temperature-dependent infrared study of bulk PtSe$_2$ and PtTe$_2$ single crystals. Using band-resolved density functional theory (DFT) calculations, we identify the bands and interband transitions responsible for the distinct optical responses of the two compounds. Because they are isostructural and isovalent, comparing them shows how the energy of the chalcogen-$p$ states relative to the Pt-$d$ states affects the free-carrier and interband responses. We also examine the contribution of transitions between the two bands forming the type-II Dirac cones.

\section{Methods}
High-quality single crystals of PtSe$_2$ and PtTe$_2$ were grown using the chemical vapor transport (CVT) method with iodine as the transport agent, following previously reported procedures~\cite{yang2018quantum,soled1975}. Initially, polycrystalline precursors were synthesized by a conventional solid-state reaction using high-purity elemental platinum (Alfa Aesar, 99.95\%), selenium (Alfa Aesar, 99.999\%), and tellurium (Alfa Aesar, 99.999\%) in their respective stoichiometric ratios. The starting materials were sealed in evacuated fused-silica ampoules and heated at 650~$^\circ$C for 120~h. The resulting polycrystalline precursors were subsequently used for single-crystal growth via CVT in the presence of iodine (Alfa Aesar, 99.998\%) at a concentration of 6~mg~cm$^{-3}$. The growth was carried out for one week under a temperature gradient of 850~$^\circ$C (source) to 750~$^\circ$C (sink) for PtSe$_2$ and 875~$^\circ$C (source) to 690~$^\circ$C (sink) for PtTe$_2$. The resulting crystals exhibited a characteristic plate-like morphology with well-defined hexagonal facets, consistent with the layered hexagonal crystal structures of PtSe$_2$ and PtTe$_2$.

Temperature-dependent reflectivity measurements were performed in the $ab$ plane over a broad frequency range from 80 to 18\,000~cm$^{-1}$ (10~meV--2.23~eV) at temperatures from 10 to 300~K. Below 600~cm$^{-1}$, a Bruker IFS~113v spectrometer was used, and the absolute value of the reflectivity was obtained by the in situ gold-overcoating technique~\cite{Homes1993}. At higher frequencies, the reflectivity was measured with a Bruker Vertex~80v spectrometer coupled to a Hyperion microscope, with a gold mirror as reference. The optical conductivity $\sigma_1(\omega)$ and the complex dielectric function $\tilde\varepsilon(\omega)$ were obtained by Kramers--Kronig analysis~\cite{DresselGruner2002}, utilizing a Hagen--Rubens extrapolation at low frequencies and x-ray scattering functions at high frequencies~\cite{Tanner2015}.

Band-structure and optical-conductivity calculations were performed in the WIEN2k code~\cite{Blaha2018wien2k,Blaha2020wien2k} using the lattice parameters of PtSe$_2$~\cite{Thomassen1929} and PtTe$_2$~\cite{Kim1999} and the Perdew--Burke--Ernzerhof (PBE) exchange-correlation functional~\cite{Perdew1996}. Spin--orbit coupling was included in all calculations. Self-consistent calculations were converged on a $14 \times 14 \times 9$ $k$ mesh, and the optical conductivity was obtained using the OPTIC module~\cite{AmbroschDraxl2006} on denser $k$ meshes of $67 \times 67 \times 43$ for PtSe$_2$ and $90 \times 90 \times 60$ for PtTe$_2$.

\begin{figure}
\centering
\includegraphics[width=0.48\textwidth]{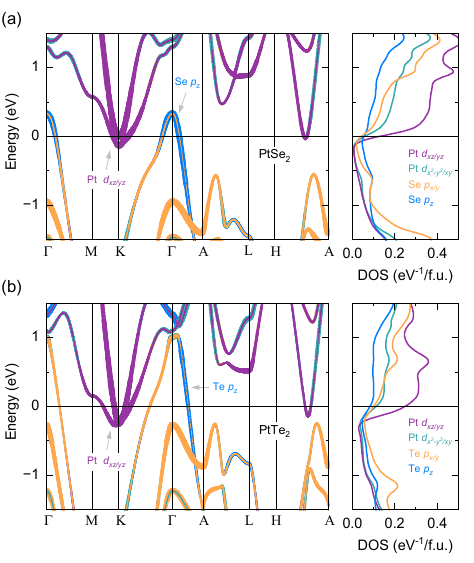}
\caption{(a) Calculated orbital-resolved band structure (left) and density of states (right) of PtSe$_2$. The line thickness represents the contributions of the Pt $d_{xz/yz}$ and $d_{x^2-y^2/xy}$ orbitals and the chalcogen $p_{x/y}$ and $p_z$ orbitals. The bands correspond to those labeled in Figs.~\ref{Fig2}(b) and~\ref{Fig2}(e). (b) Same as (a) for PtTe$_2$.}
\label{Fig3}
\end{figure}
\section{Results}
The reflectivity and the real part of the dielectric function $\varepsilon_1(\omega)$ of PtSe$_2$ and PtTe$_2$ are shown in Figs.~\ref{Fig1}(a) and~\ref{Fig1}(d), respectively. The reflectivity approaches unity in the low-frequency limit, as expected for a metal, and drops at the plasma edge. The zero crossing of $\varepsilon_1(\omega)$ corresponds to the screened plasma frequency $\omega_p^{\mathrm{scr}}$, which is 1764~cm$^{-1}$ (0.22~eV) in PtSe$_2$ and 4148~cm$^{-1}$ (0.51~eV) in PtTe$_2$ at 10~K. The screened plasma frequency can also be estimated from the frequency at which the loss function $-\mathrm{Im}[1/\tilde{\varepsilon}(\omega)]$ reaches its maximum, 1771 and 4050~cm$^{-1}$, respectively (Appendix~\ref{app:loss}).

The real part of the optical conductivity $\sigma_1(\omega)$ is presented in Figs.~\ref{Fig1}(b) and~\ref{Fig1}(e), and its decomposition within the Drude--Lorentz approach at 10~K in Figs.~\ref{Fig1}(c) and~\ref{Fig1}(f). At 10~K, the plasma frequency $\omega_p$ of the Drude term amounts to 8758~cm$^{-1}$ in PtSe$_2$ and 27\,632~cm$^{-1}$ in PtTe$_2$, corresponding to Drude spectral weights, given by the area under the Drude contribution to $\sigma_1(\omega)$, of $2.0\times10^{6}$ and $2.0\times10^{7}~\Omega^{-1}\,\mathrm{cm}^{-2}$, respectively. This spectral weight is proportional to $n/m^{*}$, where $n$ is the carrier density and $m^{*}$ the effective mass~\cite{DresselGruner2002,Tanner2019,Basov2011}. The scattering rates $\gamma$ are of the same order of magnitude, 169~cm$^{-1}$ in PtSe$_2$ and 245~cm$^{-1}$ in PtTe$_2$. The lower resistivity of PtTe$_2$ (Appendix~\ref{app:resistivity}) is therefore due to its tenfold larger spectral weight. The temperature dependence of the Drude parameters is presented in Appendix~\ref{app:drude}. 

The screened plasma frequency varies by about 2\% between 10 and 300~K. This weak temperature dependence contrasts with layered dichalcogenides in which charge-density-wave or Lifshitz transitions strongly change the free-carrier response~\cite{mathew2025interlayer,wu2015temperature}. A strong temperature dependence is also found in the semimetal ZrTe$_5$, in which band edges within a few $k_BT$ of $E_{\mathrm{F}}$ cause $\omega_p^{\mathrm{scr}}$ to vary by a factor of three~\cite{xu2018zrte5}. The screened plasma frequency is related to the Drude plasma frequency by $\omega_p = \omega_p^{\mathrm{scr}}\sqrt{\varepsilon_\infty}$, where the background dielectric constant $\varepsilon_\infty$ describes the screening of the free-carrier response by interband transitions~\cite{DresselGruner2002}. At 10~K, this relation yields $\varepsilon_\infty \approx 25$ in PtSe$_2$ and 44 in PtTe$_2$.

A narrow absorption peak appears on the tail of the Drude response. In the Drude--Lorentz fits at 10~K, this peak is centered at 1273~cm$^{-1}$ in PtSe$_2$ and at 1131~cm$^{-1}$ in PtTe$_2$ [insets of Figs.~\ref{Fig1}(c) and~\ref{Fig1}(f)]. This low-energy peak is clearly resolved only at low temperatures [insets of Figs.~\ref{Fig1}(b) and~\ref{Fig1}(e)], because the increasing scattering rate enhances the Drude contribution at the peak frequency upon warming. Above this peak, the interband responses of the two compounds are distinctly different. In PtSe$_2$, $\sigma_1(\omega)$ remains low between the low-energy peak and a strong absorption band centered at 12\,131~cm$^{-1}$ (1.50~eV). In PtTe$_2$, $\sigma_1(\omega)$ decreases above the low-energy peak along the Drude tail but remains sizable across the mid-infrared because of a broad interband absorption band. In the fit, this mid-infrared continuum is described by an additional broad oscillator centered at 2085~cm$^{-1}$. A mid-infrared absorption in a similar spectral range, centered near 2500~cm$^{-1}$, was also reported for PtTe$_2$ films~\cite{macis2024terahertz}.

Figures~\ref{Fig2}(a) and~\ref{Fig2}(d) compare the experimental interband optical conductivity at 10~K, obtained by subtracting the Drude contribution from the spectra, with the calculated in-plane interband optical conductivity. The calculations reproduce the low-energy peak, the low conductivity of PtSe$_2$ above this peak, and the continuum of PtTe$_2$. For PtSe$_2$, the energy axis of the calculated spectra is multiplied by a factor of 1.21 to match the experiment, whereas no scaling is required for PtTe$_2$. The calculated interband transitions of PtSe$_2$ thus lie at lower energies than observed, consistent with the tendency of semilocal functionals to underestimate band separations. For the same reason, the conduction bands of PtSe$_2$ had to be shifted upward by 0.18~eV to reproduce the Fermi surface measured by quantum oscillations~\cite{yang2018quantum}. We therefore treat the scaling as a phenomenological correction. It changes only the energies of the calculated features and leaves the band dispersions, orbital characters, and band assignments unchanged. The calculated out-of-plane conductivity is given in Appendix~\ref{app:anisotropy}.

We decompose the calculated interband conductivity into contributions from individual band pairs [Figs.~\ref{Fig2}(c) and~\ref{Fig2}(f)], with the bands labeled as in Figs.~\ref{Fig2}(b) and~\ref{Fig2}(e). The low-energy peak originates from C$\rightarrow$D transitions near $K$, whereas most of the interband conductivity arises from B$\rightarrow$C transitions. In PtSe$_2$, the B$\rightarrow$C transitions begin at about 2125~cm$^{-1}$ on the scaled energy axis, above the low-energy peak. This explains the weak interband conductivity between this peak and the strong absorption band at 1.50~eV. In PtTe$_2$, they already set in at about 976~cm$^{-1}$, below the low-energy peak, and form the mid-infrared continuum. The contrasting interband responses point to a different energy of band B relative to band C in the two compounds.

\section{Discussion}
Figure~\ref{Fig3} shows the orbital character of the bands labeled in Fig.~\ref{Fig2}. Bands C and D are dominated by Pt $d_{xz/yz}$ states near $E_{\mathrm{F}}$. The low-energy peak corresponds to transitions from band C into band D near $K$. In PtTe$_2$, the minimum of band C at $K$ lies about 0.1~eV lower than in PtSe$_2$, and band D also moves below $E_{\mathrm{F}}$ at $K$. The energy separation between the two bands near $K$, however, changes little, so the low-energy peak remains at a similar energy, 1273~cm$^{-1}$ in PtSe$_2$ and 1131~cm$^{-1}$ in PtTe$_2$.

Band B forms the hole pocket around $\Gamma$ identified in quantum-oscillation studies~\cite{yang2018quantum,zheng2018detailed,pavlosiuk2018galvanomagnetic}. It derives mainly from chalcogen-$p$ orbitals near $E_{\mathrm{F}}$ in both compounds. In PtTe$_2$, band B is shifted upward relative to bands C and D, and its maximum at $\Gamma$ lies about 0.7~eV higher than in PtSe$_2$. Band B acquires Pt $d_{x^2-y^2/xy}$ character close to this maximum. This change is confined to the top of the band, above the states near $E_{\mathrm{F}}$ that govern the free-carrier response and the mid-infrared continuum. This shift agrees with earlier calculations~\cite{hooda2021electronic,guo1986electronic,dai2003trends}. In these calculations, the antibonding combination of the chalcogen-$p_z$ orbitals of adjacent layers shifts upward from S to Te, because the chalcogen-$p$ level rises and the interlayer $p_z$ interaction becomes stronger. Together with the lowering of band C, this shift increases the semimetallic overlap between the maximum of band B at $\Gamma$ and the minimum of band C at $K$ from 0.46~eV in PtSe$_2$ to 1.26~eV in PtTe$_2$. This is qualitatively consistent with the larger Drude spectral weight of PtTe$_2$. The same shift brings the occupied states of band B closer in energy to the unoccupied states of band C, which lowers the onset of transitions between them and produces the continuum.

In PtTe$_2$, this continuum also damps the plasmon, which lies at a higher energy than in PtSe$_2$ and decays into interband excitations. This damping is consistent with the loss-function maximum of PtTe$_2$ being about a factor of 8 lower than that of PtSe$_2$ (Appendix~\ref{app:loss}). The plasmon of PtTe$_2$ coincides in energy with that observed by EELS~\cite{politano2018dirac,ghosh2019broadband}. It is thus accounted for by the Drude spectral weight of all bands crossing $E_{\mathrm{F}}$, screened and damped by interband transitions, without requiring an effective Dirac-cone model.

In our calculations, the type-II Dirac points lie along the $\Gamma$--$A$ direction at about $-1.27$~eV in PtSe$_2$ and $-0.80$~eV in PtTe$_2$ (Fig.~\ref{Fig3}). Bands A and B in Figs.~\ref{Fig2}(b) and~\ref{Fig2}(e) are the lower and upper branches of the Dirac cone. Near the Dirac points, both bands are fully occupied, so that A$\rightarrow$B transitions are Pauli blocked. The band-resolved decomposition shows that B$\rightarrow$C transitions dominate the calculated interband conductivity. A$\rightarrow$B transitions contribute in the energy range near 1 and 1.4~eV, consistent with the EELS excitations attributed to transitions between the two bands forming the Dirac cone~\cite{politano2018dirac}, but they remain weaker than B$\rightarrow$C transitions in $\sigma_1(\omega)$. Band B thus shapes the infrared response through its states near $E_{\mathrm{F}}$ rather than through the Dirac points.

\section{Conclusions}
We have compared the in-plane infrared response of bulk PtSe$_2$ and PtTe$_2$ single crystals from 10 to 300~K with band-resolved DFT calculations. Both compounds show a low-energy interband peak at a similar energy, arising from transitions between two Pt-$d$ bands near $K$. The differences between the two compounds are mainly due to a band of predominantly chalcogen-$p$ character, which lies higher in energy in PtTe$_2$. This upward shift increases the semimetallic band overlap, consistent with the tenfold larger Drude spectral weight of PtTe$_2$. It also lowers the onset of transitions into a Pt-$d$ band and produces the mid-infrared continuum, which damps the plasmon. The type-II Dirac points lie far below the Fermi level, where both bands forming the Dirac cones are occupied, so that they leave no signature in the infrared response.

\begin{acknowledgments}
The authors thank Maxim Wenzel and Savita Priya for fruitful discussions and Sarah Parks for preparing the samples used in the resistivity measurements. J.E.L. was supported by a Humboldt Research Fellowship from the Alexander von Humboldt Foundation. This work was supported by the National Research Foundation of Korea (NRF) grant funded by the Korea government (MSIT) (No.~RS-2024-00341731).
\end{acknowledgments}

\appendix
\section{Loss function}
\label{app:loss}
\begin{figure}[!htb]
\centering
\includegraphics[width=0.48\textwidth]{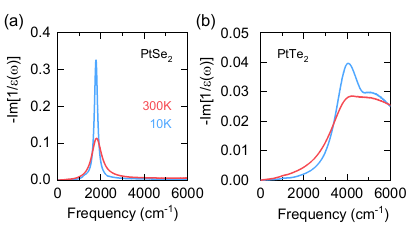}
\caption{Loss function $-\mathrm{Im}[1/\tilde{\varepsilon}(\omega)]$ of (a) PtSe$_2$ and (b) PtTe$_2$ at 10 and 300~K.}
\label{fig:lossfunction}
\end{figure}

The loss function is obtained from the complex dielectric function as $L(\omega) = -\mathrm{Im}[1/\tilde{\varepsilon}(\omega)]$. Figure~\ref{fig:lossfunction} compares the loss functions of PtSe$_2$ and PtTe$_2$ at 10 and 300~K. Both loss functions exhibit a peak close to the screened plasma frequency. In PtSe$_2$, the peak is sharp. In PtTe$_2$, the peak maximum is about a factor of 8 lower, and the peak is much broader because of interband damping. Upon warming, the peak positions remain nearly unchanged, whereas the peaks broaden.

\section{Resistivity}
\label{app:resistivity}
\begin{figure}[!htb]
\centering
\includegraphics[width=0.48\textwidth]{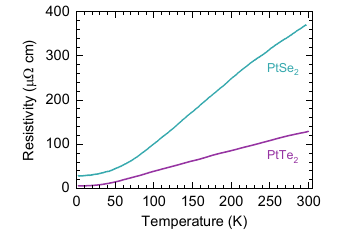}
\caption{Temperature dependence of the in-plane resistivity of PtSe$_2$ and PtTe$_2$.}
\label{fig:resistivity}
\end{figure}

Four-point resistivity measurements were performed in the $ab$ plane between 2 and 300~K. Figure~\ref{fig:resistivity} shows that both compounds exhibit metallic behavior and that the resistivity of PtTe$_2$ is lower than that of PtSe$_2$.

\section{Temperature dependence of the Drude parameters}
\label{app:drude}
\begin{figure}[!htb]
\centering
\includegraphics[width=0.48\textwidth]{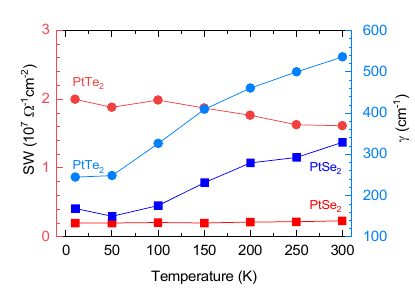}
\caption{Temperature dependence of the Drude spectral weight (red, left axis) and the scattering rate $\gamma$ (blue, right axis) obtained from the Drude--Lorentz fits. Circles and squares denote PtTe$_2$ and PtSe$_2$, respectively.}
\label{fig:plasma}
\end{figure}

Figure~\ref{fig:plasma} shows the temperature dependence of the Drude spectral weight and the scattering rate obtained from the Drude--Lorentz fits. At all temperatures, the spectral weight of PtTe$_2$ is seven to ten times larger than that of PtSe$_2$, whereas the scattering rates differ by less than a factor of 2. In both compounds, the scattering rate approximately doubles between 10 and 300~K, from 169 to 330~cm$^{-1}$ in PtSe$_2$ and from 245 to 536~cm$^{-1}$ in PtTe$_2$. 

\section{Calculated optical anisotropy}
\label{app:anisotropy}
\begin{figure}[!htb]
\centering
\includegraphics[width=0.48\textwidth]{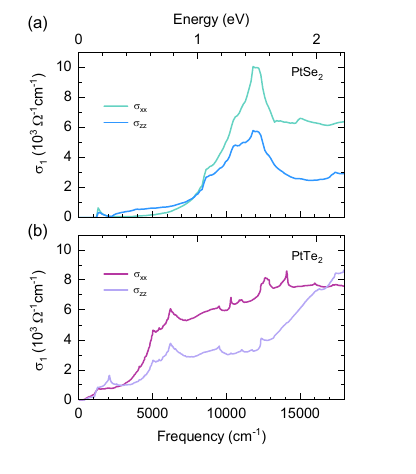}
\caption{Calculated in-plane ($\sigma_{xx}$) and out-of-plane ($\sigma_{zz}$) interband optical conductivity of (a) PtSe$_2$ and (b) PtTe$_2$.}
\label{fig:anisotropy}
\end{figure}

Figure~\ref{fig:anisotropy} compares the calculated in-plane and out-of-plane interband optical conductivities, $\sigma_{xx}(\omega)$ and $\sigma_{zz}(\omega)$, of PtSe$_2$ and PtTe$_2$. Because the reflectivity was measured on the $ab$ plane, the comparison with the experiment in the main text is based on $\sigma_{xx}(\omega)$. The calculated anisotropy is weak in both compounds.

\makeatletter
\def\bibsection{\par\begingroup\baselineskip26\p@\bib@device{\hsize}{72\p@}\endgroup\nobreak\@nobreaktrue\addvspace{19\p@}}
\makeatother
\FloatBarrier 
\bibliography{PRB}

\end{document}